\documentclass[twocolumn,preprintnumbers,amsmath,amssymb]{revtex4-2}
\usepackage{graphicx}
\usepackage{color}
\usepackage{caption}
\usepackage{subfigure}
\begin{document}

\title{Wetting as an emergent property of water: reformulating Young’s equation on molecular grounds}

\author{Nicol\'as A. Loubet$^1$}
\author{Gustavo A. Appignanesi$^{1 *}$}

\newcommand{\angstrom}{\mbox{\normalfont\AA}}

\affiliation{
$^1$ INQUISUR, Departamento de Qu\'{i}mica, Universidad Nacional del Sur (UNS)-CONICET, Avenida Alem 1253, 8000 Bah\'{i}a Blanca, Argentina\\
* Corresponding author: appignan@criba.edu.ar\\
}

\date{\today}







\begin{abstract}
Young’s equation provides a remarkably successful macroscopic description of wetting, yet its molecular origin—particularly for water—has remained elusive for over two centuries. Here we make the molecular basis of aqueous wetting explicit by reformulating it in terms of a molecular wetting coefficient, $\omega_m$, which quantifies how an interface compensates the intrinsic energetic cost of hydrogen-bond defects relative to bulk water. Across a broad and continuous spectrum of hydrophilicities, spanning chemically diverse experimental and model surfaces, macroscopic contact angles collapse onto a single universal master curve when expressed through $\omega_m$. This molecular reformulation closes Young’s and Young–Dupré relations on energetic grounds, establishing a unified and predictive physical link between wetting, adhesion, cavitation, and nanoconfined filling. By anchoring interfacial behavior to water’s intrinsic hydrogen-bond energetic scales, our results reveal wetting as an emergent property of water itself, rather than a surface-specific attribute and provide a transferable molecular framework that recalibrates energetic intuition and guides the rational design of aqueous interfaces.  (This document is the unedited Author’s version of a Submitted Manuscript subsequently accepted for publication in J. Am. Chem. Soc.; for the published version, which includes a more complete molecular-thermodynamics grounding of the method see: doi.org/10.1021/jacs.6c01106)

\end{abstract}

\maketitle

\section*{Introduction}

Wetting phenomena play a central role across a wide range of physical, chemical, and biological contexts, from capillarity and adhesion to interfacial stability, confinement, and nanoscale transport \cite{Young1805,Dupre1869,Bonn2009,deGennes1985,Israelachvili2011,Li,Giacomello2016PNAS,Si2018}. Traditionally, wetting is formulated as a macroscopic property of materials, described in terms of interfacial free energies and contact angles. Within this framework, equilibrium wetting arises from a balance between solid--vapor, solid--liquid, and liquid--vapor interfacial energies, encapsulated by Young’s equation \cite{Young1805}, and its energetic extension through the Young--Dupré relation \cite{Dupre1869}. Despite its success, this description remains phenomenological: it does not specify the molecular mechanisms that ultimately determine wetting, nor does it explain why chemically and structurally distinct surfaces often display remarkably similar wetting behavior \cite{Bonn2009}.

In aqueous environments, wetting is closely intertwined with the notions of hydrophobicity and hydrophilicity, which are widely used to rationalize the affinity of surfaces, solutes, and confined regions for water across disciplines ranging from materials science to biophysics \cite{Chandler2005,Giovambattista2012,RegoPatel2022}. Hydrophobicity has been shown to emerge from collective density fluctuations spanning molecular to mesoscopic length scales, as formalized by the Lum--Chandler--Weeks framework \cite{Lum1999}, with direct implications for wetting, cavitation, and nanoconfined filling. Yet, in practice, hydrophilicity and hydrophobicity are most often invoked qualitatively, frequently framed in terms of the presence or absence of specific surface--water interactions. While useful as descriptors, such classifications do not identify the energetic scales that govern how water responds to an interface \cite{Dill2008}.

This limitation—namely, the absence of a molecular-level energetic rationalization underlying macroscopic wetting—is particularly acute for water. Liquid water is governed by a fluctuating hydrogen-bond (HB) network, in which missing or distorted bonds carry a finite energetic penalty \cite{Luzar1996}. Nevertheless, aqueous wetting is still commonly discussed using qualitative notions of ``strong'' or ``weak'' surface--water interactions, often implicitly benchmarked against the energy of a full hydrogen bond \cite{Israelachvili2011}. Such heuristics obscure a central fact: water itself possesses intrinsic energetic preferences that largely govern its interfacial behavior, rooted in the energetic cost associated with hydrogen-bond defects within the fluctuating network \cite{Stillinger1980,Chandler2005}.

Recent work has shown that bulk water exhibits robust, model- and temperature-independent energetic thresholds associated with the compensation of hydrogen-bond defects \cite{Appignanesi2024}. In particular, the energetic cost of a missing hydrogen bond is partially compensated at a characteristic defect interaction threshold (DIT $\approx -6$~kJ~mol$^{-1}$)\cite{Appignanesi2024}, while approximately twice this value ($\approx -12$~kJ~mol$^{-1}$) separates defective from tetrahedral local environments \cite{Appignanesi2025,Appignanesi2020}. These thresholds imply that partial hydrophilicity corresponds to compensating only $\sim 30\%$ of a typical hydrogen-bond energy, and strong hydrophilicity to $\sim 60\%$, fundamentally recalibrating long-standing energetic intuitions about water--surface interactions.

The central question we address here is how these intrinsic energetic properties of water manifest in macroscopic wetting and interfacial thermodynamics. We show that aqueous wetting can be reformulated in terms of a molecular wetting coefficient, $\omega_m$, derived from a microscopic energetic descriptor of interfacial water. When expressed through $\omega_m$, contact angles across chemically diverse systems collapse onto a single universal master curve, and wetting, adhesion, cavitation, and nanoconfined filling are unified within a single molecular framework. This perspective reveals wetting as an emergent property of water itself, rather than a surface-specific attribute, and establishes a quantitative bridge between molecular energetics and macroscopic wetting behavior. More generally, by anchoring interfacial phenomena to intrinsic energetic scales of water, this framework provides a transferable basis for understanding and rationally controlling aqueous behavior across diverse interfacial and confinement scenarios.

\section*{Molecular reformulation of Young’s equation}

At macroscopic scales, equilibrium wetting is described by Young’s equation,
\begin{equation}
\gamma_{sv} = \gamma_{sl} + \gamma_{lv} \cos\theta ,
\end{equation}
which expresses the balance of interfacial free energies at the three-phase contact line. Rearranging, one obtains
\begin{equation}
\cos\theta = \frac{\gamma_{sv} - \gamma_{sl}}{\gamma_{lv}} .
\end{equation}

While this expression successfully predicts contact angles, it leaves open a fundamental question: what microscopic physics determines the difference $\gamma_{sv} - \gamma_{sl}$ for water? In particular, it does not specify how the energetic preferences of liquid water —dominated by its hydrogen-bond network— enter this balance.

For aqueous interfaces, wetting can be understood as a local thermodynamic competition governed by hydrogen-bond defects of interfacial water. Contact with a surface necessarily perturbs the bulk hydrogen-bond network, generating missing or distorted bonds that carry a finite energetic cost. Whether wetting occurs therefore depends on the ability of the interface to compensate this defect penalty. This observation motivates a molecular interpretation of the Young balance, in which $\gamma_{sv} - \gamma_{sl}$ is no longer treated as an abstract material parameter, but rather as a measure of the energetic stabilization experienced by interfacial water relative to bulk liquid water.

\section*{Molecular wetting coefficient}

Bulk water exhibits a well-defined energetic scale associated with hydrogen-bond defects. The defect interaction threshold (DIT) marks the energetic compensation required to stabilize a missing hydrogen bond. Interfacial water molecules stabilized less favorably than this threshold behave, in an energetic sense, as defective bulk water.

This stabilization is quantified\cite{Appignanesi2024} through a microscopic interaction measure, $V_{4S}$, defined as the interaction energy of the weakest one among the four tetrahedral interacting sites of a water molecule (where each site includes both water-water and water-surface interactions). Normalizing this quantity by the intrinsic defect scale of bulk water leads naturally to a dimensionless molecular wetting coefficient,
\begin{equation}
\omega_m = \frac{\langle V_{4S} \rangle}{\mathrm{DIT}} - 1 .
\end{equation}

By construction, $\omega_m = 0$ marks the point at which the energetic penalty of interfacial hydrogen-bond defects is exactly compensated. Negative values correspond to interfaces that fail to stabilize interfacial water (hydrophobic regime), while positive values indicate net stabilization (hydrophilic regime).

Crucially, $\omega_m$ is defined entirely in terms of water’s intrinsic energetic scales. Although evaluated from interfacial configurations, it does not encode surface chemistry explicitly; instead, it measures how effectively a surface couples to the hydrogen-bond network of water. To a very good approximation, the Young balance can thus be recast in terms of $\omega_m$ as
\begin{equation}
\cos\theta \equiv \omega_m .
\end{equation}
In this form, Young’s equation acquires a direct molecular interpretation: the contact angle reflects whether, and to what extent, the surface compensates the energetic cost of hydrogen-bond defects of interfacial water.

\section*{Universal wetting behavior}

An immediate consequence of this reformulation is that chemically diverse surfaces should exhibit similar wetting behavior when characterized by $\omega_m$. Indeed, once contact angles are expressed in molecular terms, surface-specific details become secondary to the universal energetic preferences of water.

\begin{figure}[t]
\centering
\includegraphics[width=0.5\textwidth]{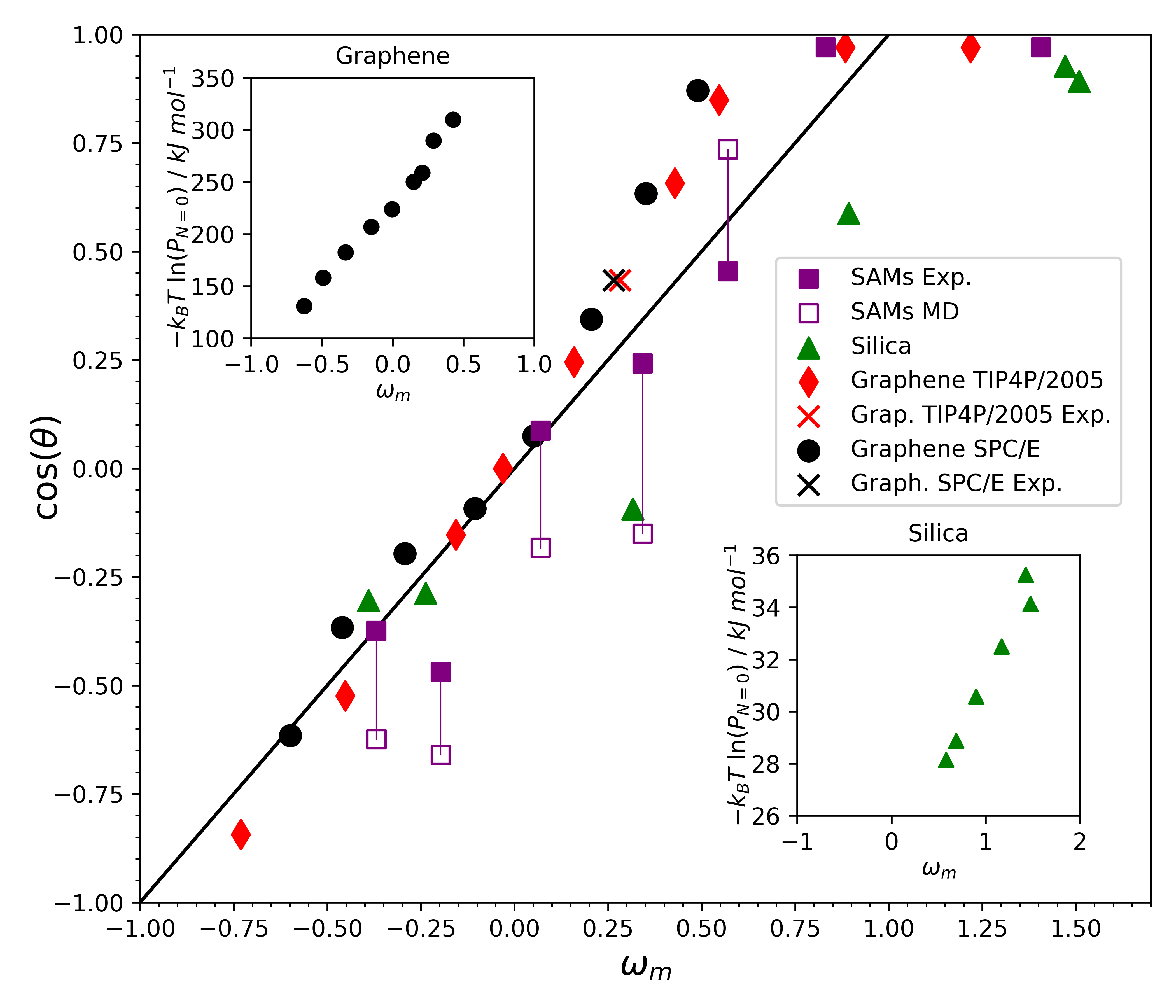}
\caption{\small{\textbf{Universal molecular description of aqueous wetting.}
Contact angles for chemically diverse systems plotted as a function of the molecular wetting coefficient $\omega_m$.  Experimental data for self-assembled monolayers (SAMs) with different functionalization (from apolar to polar groups), together with results for dispersive graphene (with variable Lennard--Jones attractions) and polar silica (with varying polarity),  collapse onto a single master curve defined by $\cos\theta=\omega_m$.
Inset: Cavitation free energy for a small (methane-like sized) sphere tangent to the surface, $-k_BT\ln P(N=0)$, plotted against $\omega_m$ for graphene and silica, demonstrating a thermodynamic closure linking wetting, adhesion, and cavitation. (Contact angle data for simulated SAMs, experimental SAMS, SPC/E on graphene, TIP4P/2005 on graphene, silica and experimental graphene were extracted from \cite{Godawat,Sigal,Werder,Wloch,Giovambattista,Li}, respectively; for simulation details and calculation of the molecular wetting coefficient, please refer to the Supplementary material).
}}
\label{fig:fig1}
\end{figure}

Figure~\ref{fig:fig1} shows that contact angles for experimental self-assembled monolayers (SAMs), dispersive graphene, and polar silica collapse onto a single master curve defined by $\cos\theta=\omega_m$. The systems span a wide range of surface chemistries and interaction mechanisms, yet follow the same molecular relation.

Graphene plays a particularly instructive role. As a purely dispersive surface incapable of forming hydrogen bonds, graphene nevertheless follows the same molecular wetting relation as polar substrates. This demonstrates that aqueous wetting is not controlled by the formation of specific surface–water hydrogen bonds, but by the degree to which the interface stabilizes water relative to its bulk hydrogen-bond network.

Experimental data display remarkably small scatter across chemically distinct systems (compare for example the results for experimental vs simulated SAMs). Molecular-dynamics results show larger deviations in the strongly hydrophilic regime, which arise from known methodological limitations of nanodroplet simulations—finite-size effects, line tension, and enhanced interfacial structuring—rather than from a breakdown of the molecular framework.

\section*{Young--Dupré relation, adhesion, and cavitation}

Molecular simulations have shown that wetting, adhesion and cavitation phenomena at aqueous interfaces and under nanoconfinement are governed by subtle free-energy balances and collective density fluctuations, rather than by local surface chemistry alone \cite{Patel2010Fluctuations,Giacomello2017PNAS,RegoPatel2022}. However, a unified molecular criterion directly linking these phenomena to macroscopic wetting relations has so far been lacking.

The molecular reformulation introduced here provides such a closure. Specifically, the Young--Dupré relation defines the work of adhesion as
\begin{equation}
W_{\mathrm{adh}} = \gamma_{lv} (1 + \cos \theta) ,
\end{equation}
which, in terms of the molecular wetting coefficient, gives
\begin{equation}
W_{\mathrm{adh}} = \gamma_{lv} (1 + \omega_m) .
\end{equation}

This expression shows that the work of adhesion depends linearly on $\omega_m$, directly linking a macroscopic interfacial quantity to the microscopic energetic coupling between the surface and water’s hydrogen-bond network. This linear dependence indicates that $\omega_m$ not only captures macroscopic wetting and adhesion, but also encodes the energetic balance underlying interfacial density fluctuations. The free energy required to create an empty cavity near a surface is given by
\begin{equation}
\Delta G_{\mathrm{cav}} \propto -k_B T \ln P(N=0),
\end{equation}
where $P(N=0)$ is the probability of finding zero water molecules in a defined observation volume \cite{Godawat,Patel2010Fluctuations,RegoPatel2022}.
As shown in the inset of Fig.~1, $-k_B T \ln P(N=0)$ varies linearly with $\omega_m$ for both graphene and silica, confirming a thermodynamic closure linking wetting, adhesion and cavitation.

\section*{Nanoconfinement and thermodynamic closure}

If $\omega_m$ captures the balance between interfacial stabilization and hydrogen-bond defect formation at a free surface, the same balance should become even more stringent under geometric confinement. Nanoconfinement therefore provides a natural arena to test the predictive power of the molecular wetting coefficient, as filling and cavitation amplify the energetic cost of defect formation.

\begin{figure}[t]
\centering
\includegraphics[width=0.5\textwidth]{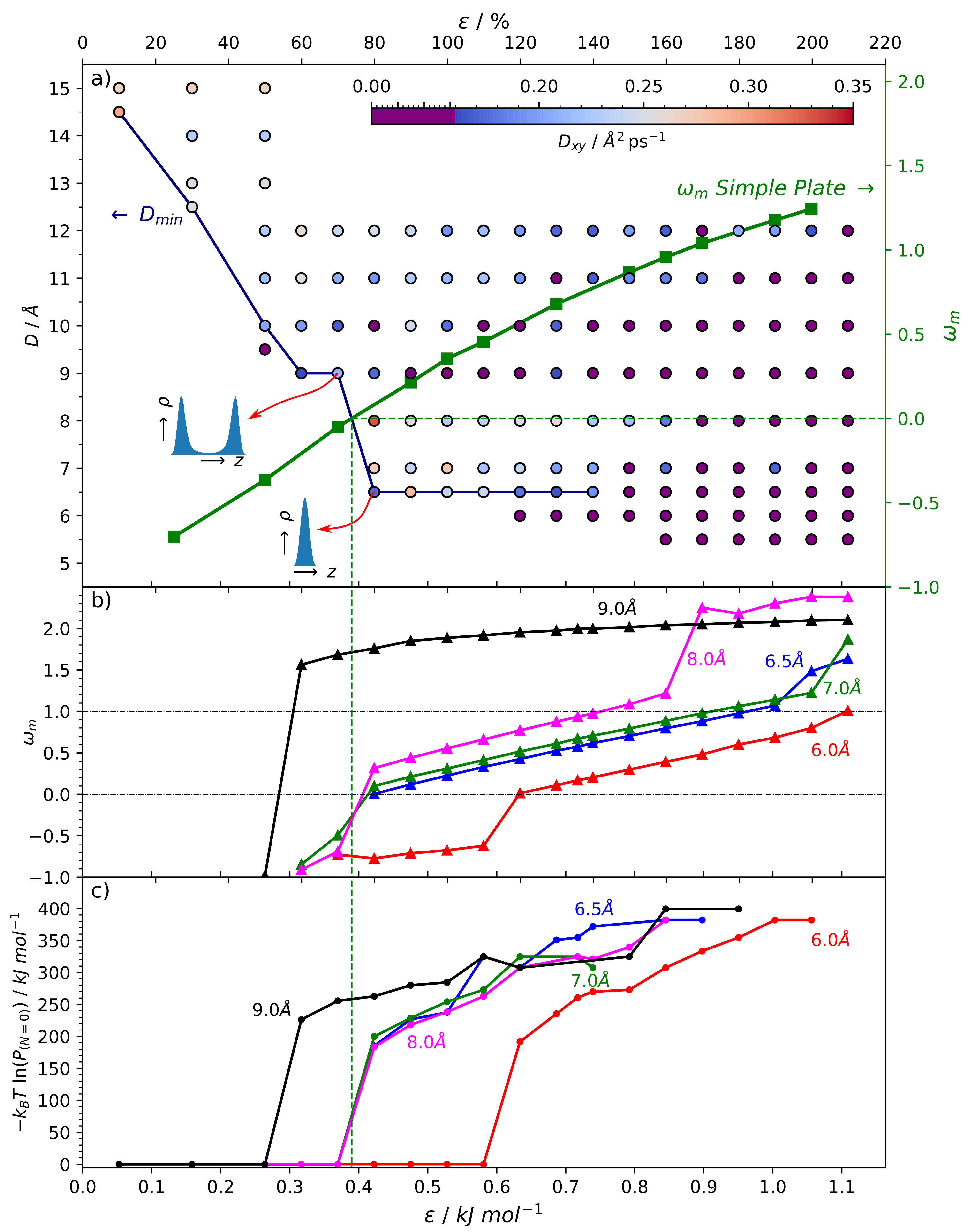}
\caption{\small{\textbf{Nanoconfinement, filling, and thermodynamic closure controlled by the molecular wetting coefficient.}
(a) Minimum stable plate separation $D_{\min}$ for water confined between two parallel graphene sheets as a function of interaction strength $\varepsilon$. Water density profiles for 70\%--9~\AA\ and 80\%--6.5~\AA\ are sketched as insets. The green curve reports $\omega_m$ for a single graphene surface; the vertical dashed green line (that extends down to the other two panels of the figure) marks the value of $\varepsilon$ for which  $\omega_m=0$. The color of the symbols indicates the in-plane (2D) diffusion coefficient of confined water, from more mobile (red) to nearly arrested or frozen configurations (violet).
(b) Molecular wetting coefficient within the confined region for different separations.
(c) Cavitation free energy $-k_BT\ln P(N=0)$ within the confined region for the same confinements.}}
\label{fig:fig2}
\end{figure}

Figure~\ref{fig:fig2}a reports the minimum plate separation $D_{\min}$ at which water can stably exist between two parallel graphene sheets as a function of the solid–water interaction strength $\varepsilon$. At low interaction strengths, water is only stable at relatively large separations ($D \gtrsim 8$–9~\AA), where confinement is weak and the confined region fills in a multilayer fashion (see the sketched water density profile for 70\%--9~\AA). In this regime, water can sustain a hydrogen-bond network that remains largely bulk-like, and the energetic penalty associated with confinement-induced defects is only weakly amplified.

Upon increasing $\varepsilon$, a sharp transition is observed: the confined region suddenly fills down to $D_{\min}\approx 6.5$~\AA, which corresponds to the steric limit for liquid water between graphitic walls and signals the formation of a strictly two-dimensional monolayer, as sketched for 80\%--6.5~\AA\ (this liquid monolayer regime continues for larger water-wall interactions until it freezes, as depicted by the 2D diffusion coefficient). This transition is abrupt rather than progressive, indicating a genuine thermodynamic crossover between a weakly confined multilayer state and a stabilized 2D phase. The green curve in Fig.~\ref{fig:fig2}a reports the molecular wetting coefficient $\omega_m$ for a single graphene surface, showing that this confinement transition coincides with the crossing of $\omega_m=0$.

This correspondence becomes more explicit in Fig.~\ref{fig:fig2}b, which shows $\omega_m$ within the confined region as a function of $\varepsilon$ for different fixed plate separations. For monolayer and near-monolayer confinements (6.5, 7.0, and 8.0~\AA), $\omega_m$ exhibits a sharp increase precisely at the interaction strength where $\omega_m$ for a single plate crosses zero. This indicates that stable monolayer confinement is only achieved once each wall independently compensates the hydrogen-bond defect cost imposed by confinement. In contrast, for larger separations (9.0\AA), the increase in $\omega_m$ occurs at substantially lower $\varepsilon$, reflecting the fact that multilayer water can form a more bulk-like hydrogen-bond network and therefore requires less interfacial stabilization.

Figure~\ref{fig:fig2}c provides the thermodynamic closure by reporting the cavitation free energy, $-k_BT\ln P(N=0)$, within the interplate region, for the same confinements. For monolayer-like separations, the cavitation cost remains negligible until $\varepsilon$ reaches the value at which $\omega_m$ of a single plate becomes positive, after which it rises sharply. This demonstrates that suppression of cavitation, stabilization of a dense confined phase, and the emergence of positive $\omega_m$ are tightly linked. For the largest separation (9.0\AA), cavitation costs increase at lower $\varepsilon$, consistent with the earlier stabilization of a bulk-like multilayer liquid. At the smallest separation (6.0~\AA), although large interaction strengths are required, the confined water does not enter a liquid state but instead adopts a highly ordered, frozen-like configuration.

Together, these results show that nanoconfinement does not introduce new wetting physics but amplifies the same molecular criterion already present at a free interface. Aligning the three panels of Figure~\ref{fig:fig2} reveals that the filling, cavitation, and confinement transitions all occur when the molecular wetting coefficient of a single surface crosses $\omega_m=0$ (see vertical green dotted line that spans the three panels of Figure~\ref{fig:fig2}). Figure~\ref{fig:fig2} also hints at another physical lesson: greater attraction is not always better. Under strong confinement (as in 6.5~\AA), increasing surface–water interactions can suppress mobility and promote solid-like ordering rather than enhancing fluidity. Something similar (though milder) is suggested by the relative saturation of $\omega_m$ and the increase in stiffness reflected in the cavitation free energy.

\section*{Conclusions}

We have shown that wetting can be understood as an emergent property of water itself. The molecular wetting coefficient $\omega_m$ provides a unified microscopic descriptor of contact angles, adhesion, cavitation, and nanoconfined filling across chemically diverse systems. By reformulating Young’s and Young–Dupré relations on molecular grounds, this framework establishes a direct and quantitative bridge between classical wetting theory and the microscopic energetics of liquid water.

In doing so, it closes a conceptual loop opened more than two centuries ago, when Thomas Young related surface tension to molecular cohesion using water itself. Our results demonstrate that the same intrinsic energetic preferences of water continue to govern wetting behavior across interfaces, chemistries, and length scales.

Beyond providing a molecular closure of Young’s equation, this framework explicitly recalibrates long-standing energetic intuitions about hydrophilicity. We find that relatively modest surface–water interactions—well below the energy of a full hydrogen bond—are sufficient to promote wetting, filling, and cavitation suppression. Conversely, increasing surface attraction beyond this intrinsic scale does not necessarily improve interfacial behavior and may even be detrimental under strong confinement, where excessive stabilization leads to reduced mobility and solid-like ordering. Wetting, therefore, is not optimized by maximizing attraction, but by matching water’s intrinsic energetic scales.

More broadly, by identifying the intrinsic energetic thresholds that control aqueous interfacial behavior, this work shifts the focus from surface-specific chemistry to water-centered rational design principles. It highlights weakly to moderately hydrophilic regimes—and functional groups spanning this region of the hydrophilicity spectrum—as effective and tunable handles for controlling aqueous interfaces across a wide range of physical and chemical contexts.

\section*{Computational Methods}

All results reported in this work are based on classical molecular dynamics simulations of water at ambient conditions in contact with chemically and structurally diverse interfaces, including self-assembled monolayers with different functionalizations ($SAM-CH_3$, $SAM-CF_3$, $SAM-OCH_3$, $SAM-CONHCH_3$, $SAM-CH_2CN$, $SAM-CONH_2$, and $SAM-OH$), hydroxylated silica surfaces with tunable polarity, and graphene-based model systems with systematically varied dispersive interactions.

Simulations were performed in the isothermal--isobaric ensemble using established force-field models for water and solid substrates. Interfacial wetting behavior was characterized through macroscopic contact angles, nanoconfined filling and cavitation, and microscopic structural and thermodynamic descriptors of interfacial water.

The molecular wetting coefficient $\omega_m$ was obtained from the structural indicator $V_{4S}$ \cite{Appignanesi2024} evaluated on inherent structures of interfacial water, providing a dimensionless measure of how an interface compensates the energetic cost of hydrogen-bond defects relative to bulk water. As already indicated, full details are provided in the Supplementary Material. In simple terms, $V_{4S}$ is given by the weakest of the four tetrahedrally-arranged interaction sites of the water molecule (incorporating both water-water and water-surface interactions). In turn, the  dimensionless molecular wetting coefficient is here defined by $\omega_m = \frac{\langle V_{4S} \rangle}{\mathrm{DIT}} - 1$.

Cavitation propensities were quantified through density fluctuation statistics based on the probability of forming empty probe volumes near the interface.

Full details of systems construction, force-field parameters, simulation protocols, complete definitions of microscopic observables, and statistical analysis procedures are provided in the Supplementary Material (SM.pdf file).

\section*{Data Availability}

A dataset including fully reproducible starting points for the MD
simulations of water in contact with chemically and structurally diverse interfaces, as well as under nanoconfinement, are provided in https://zenodo.org/records/18330308. All systems and protocols closely follow previously published and validated studies, as detailed in the Supplementary Material. For space reasons, only initial configurations and input files are provided. Full production trajectories are available from the corresponding author upon reasonable request.

\end{document}